# Crowd turbulence: the physics of crowd disasters

Dirk Helbing[1,2],   Anders Johansson[1],   HE Habib Z. Al-Abideen[3]

(1. Institute for Transport & Economics, Dresden University of Technology, D-01062 Dresden, Germany;
2. Collegium Budapest Institute for Advanced Study, Szentháromság u. 2, 1014 Budapest, Hungary;
3. The Central Directorate for Holy Areas Development, Municipal and Rural Affairs,
Riyadh, Kingdom of Saudi Arabia)

**Abstract**  The panic stampede is a serious concern during mass events like soccer championship games. Despite huge numbers of security forces and crowd control measures, hundreds of lives are lost in crowd disasters each year. An analysis of video recordings of the annual pilgrimage in Makkah reveals how high-density crowds develop to turbulent dynamics and earthquake-like eruptions, which is impossible to control.

## 1  Introduction

Unidirectional pedestrian flows are mostly assumed to move smoothly according to the "fluid-dynamic" flow-density relationship $q(\rho) = \rho v(\rho)$, where $q$ represents the flow, $\rho$ is the pedestrian density, and the average velocity $v$ is believed to go to zero at some maximum density as in traffic jams. This formula is often used as basis for the design of pedestrian facilities and evacuation analyses, although reliable empirical data of conditions potentially leading to crowd disasters are lacking.

The above description of high-density crowds and conventionally used fundamental diagrams $q(\rho)$ are absolutely misleading. Even the most accurate descriptions today[1,2] are to be corrected. We have studied videos recorded at the entrance to the Jamarat Bridge on the 12th day of the Muslim pilgrimage in 2006, when several hundred thousand pilgrims were waiting to start the stoning ritual.

## 2  Results

New video analysis techniques reveal that even at the highest levels of crowding, the local density is not homogeneous. For example, at average densities of 6 persons per m$^2$ the local densities can reach values up to 9 per m$^2$ and more, but also at extreme densities pedestrians keep moving (the average speed at 6 per m$^2$ is about 18 m/min), see Fig. 1a. The constant speed at local densities $\geq 7$ per m$^2$ indicates a collective crowd motion, in which individual control is partially lost, when pedestrian bodies are pressed against each other. It also explains an unexpected, second increase of the flow, which can cause forward and backward compression waves.

In fact, when the maximum density reaches about 7 persons per m$^2$, one can observe a surprising transition from laminar to longitudinally unstable flows, namely upstream moving stop-and-go waves[3] of period 45 s that can last for 20 minutes. A subsequent, transversal instability generates even turbulent like flows (Fig.1(b)), characterized by random displacements of pedestrians into all possible directions up to 12 meters or more. With a certain probability, large displacements cause people to fall and to be trampled. The area of trampled people spreads in the course of time.

We suggest to compare extreme crowding with driven granular media. These may form density waves, while moving forward[4]. However, under quasi-static conditions[5], force chains[6] are building up, causing strong variations in the strengths and directions of local forces. As in earthquakes,[7,8] this can lead to events of sudden, uncontrollable stress release with power-law distributed displacements. Such a power-law has in fact been discovered by our video-based crowd analysis (Fig.1(c)).

Note that in contrast to granular particles, pedestrians are self-propelled, with a propulsion force that increases in high-density areas to gain more space. In "crowd panic" people push each other to escape the pressure. Together with the compressibility of bodies and open boundary conditions, this deteriorates the situation and generates particularly violent displacements.

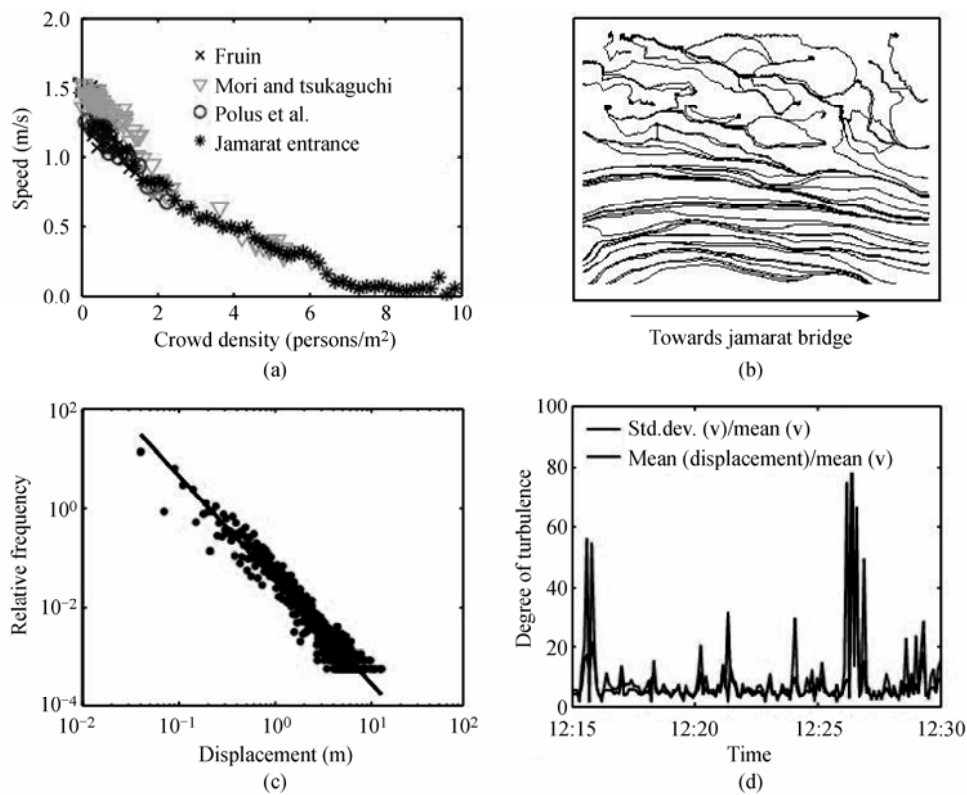

**Fig.1**  Results of our data analysis based on video recordings by fixed cameras mounted on high poles

Crowd turbulence is potentially life-threatening and experienced in hundreds of crowd-intensive events each year. Determining warning signs in advance of crowd disasters (Fig.1(d)) will enable anticipative crowd control measures. This can significantly enhance crowd safety in the future.

More than 1,000 pedestrians were identified automatically and simultaneously tracked. The local density $\rho(\boldsymbol{r},t)$ around a location $\boldsymbol{r}$ was determined via $\rho(\boldsymbol{r},t) = \frac{\beta}{\pi} \sum_i \exp[-\beta\|\boldsymbol{r}_i(t) - \boldsymbol{r}\|^2]$, where $\boldsymbol{r}_i(t)$ are the positions of the pedestrians $i$ at time $t$ and $\beta = 0.5$. This method is particularly suited for high densities. Local speeds $v$ and displacements $d$ between stopping events were determined from individual pedestrian trajectories. (a) Average local speed $v$ versus local crowd density $\rho$ at the entrance of the Jamarat Bridge in comparison with previous measurements reported in the literature. Note that the average velocity at extreme densities stays finite due to collective motion. (b) Representative pedestrian trajectories indicate that the lower area still shows transversally stable flow, while the flow in the upper part (towards the middle of the Jamarat Bridge) is already turbulent. For video films see the supplementary information. (c) Double-logarithmic representation of the frequency of differently sized displacements between stopping events. The scaling exponent of this power law is $\alpha = 2.01 \pm 0.15$. (d) The sad crowd accident on January 12, 2006, started at 12:28, directly after the time-dependent average relative displacement and the relative standard deviation of speeds reached their highest values (12:27), about 11 minutes after the first warning signs occurred (12:16).